# A LoRaWAN Wireless Sensor Network for Data Center Temperature Monitoring

Tommaso Polonelli[1], Davide Brunelli[2], Andrea Bartolini[1], Luca Benini[1,3]

[1] University of Bologna, Italy
{tommaso.polonelli2, andrea.bartolini, luca.benini}@unibo.it
[2] University of Trento, Italy - davide.brunelli@unitn.it
[3] ETH Zurich, Switzerland - luca.benini@iis.ee.ethz.ch

**Abstract.** High-performance computing installations, which are at the basis of web and cloud servers as well as supercomputers, are constrained by two main conflicting requirements: IT power consumption generated by the computing nodes and the heat that must be removed to avoid thermal hazards. In the worst cases, up to 60% of the energy consumed in a data center is used for cooling, often related to an over-designed cooling system. We propose a low-cost and battery-supplied wireless sensor network (WSN) for fine-grained, flexible and long-term data center temperature monitoring. The WSN has been operational collecting more than six million data points, with no losses, for six months without battery recharges. Our work reaches a 300x better energy efficiency than the previously reported WSNs for similar scenarios and on a 7x wider area. The data collected by the network can be used to optimize cooling effort while avoiding dangerous hot spots.

## 1 Introduction

Due to the extraordinary fast growth of IT (Information Technology) industry, the worldwide data centers energy consumption has attracted global attention because of their impact on pollution and climate change. In past years the overall trend in hardware design field continue to decrease the power efficiency of both processor and memory, but a parallel trend at data centers is that the heat density of computing systems has increased at a faster rate. Nowadays, into the data center hosting facilities, the IT devices count only for 30% to 60% [1] of the overall electric bill; indeed the rest of the energy is lost by the environmental control systems such as Computer Air Conditioning (CRAC) units, water chillers, humidifiers or during the power conversion process.

Real-time data about the temperature inside a data center building [2], together with historical information collected, are useful not only for diagnosis problems, to predict possible thermal issues, but also for improving the data center power efficiency [3]. Tier0 Supercomputing centers [4] are designed for peak computational performance and thus are characterized by the highest power/computational densities between data centers. Cooling efficiency in this domain limits the maximum achievable performance and thus Tier0's QoS. As an example, the former most powerful supercomputer worldwide Tianhe-2 occupies 720m$^2$, consumes 17.8 MW for 33.2 Petaflops. However, the power consumption increases to 24 MW also considering the cooling infrastructure [5]. This cost can be reduced by adopting predictive control approaches [6][7], which combine the supercomputer's IT power consumption, external and room temperatures for optimal control of cooling effort [8]. Traditional commercial solutions for temperature monitoring use wired sensors, but due to the high installation costs, systems utilize only a few measurements points [3]. On the other hand, wireless sensor networks (WSN) are ideal for scattered sensing systems; mobile nodes may be placed freely in critical areas to measure temperature or power consumption [9][10]. Unfortunately, the data center environment is characterized by a typical condition that is generally adverse for wireless communication. The primary material in data center facilities is metal, in addition to switches, racks, cables, and other obstacles including cooling fans, power distribution system and cable rails that generate intense electromagnetic noise.

This paper proposes a WSN designed to monitor in a distributed fashion the temperature evolution in a data center with the goal to improve cooling efficiency. We use Semtech's LoRa (Long Range) [11] that is one of the most promising wide-area IoT communication technologies [12][13] because of the proprietary spread spectrum modulation.

So far, indoor monitoring applications used mostly Zigbee and other mesh-oriented protocols in the 2,4GHz bandwidth studied intensively during the past decade. In this case, multi-hop communication is necessary for long distance transmission, or for reliability in noisy or crowded environments. The usage of LoRa in indoor environments introduces a return to the one-hop communication model, at the cost of a reduced available bandwidth, but with the capability to cover up 34000m$^2$ [14] in a single communication with a similar transmission power. Moreover, for low traffic intensity, it has been demonstrated in [15] that indoor LoRa communication is more energy efficient than a multi-hop network that needs more than one router to cover the same distance. That article shows that even the 802.11ah is significantly worse than LoRa, regarding energy efficiency, when an application requires to exchange tiny data packets.

Considering these recent results and the requirements for data-center environmental monitoring, in this paper, we describe the first deployment of a distributed temperature monitoring system in a data-center using the LoRa technology. The contribution of the paper is twofold: (i) to explore the performance of the application using LoRaWAN, especially regarding throughput, energy consumption and packet conflicts; (ii) to investigate the merits of using LoRaWAN as an alternative for the mesh-oriented communications used in this class of applications, so far.

The sensor nodes are designed for easy deployment within an existing supercomputing center facility. Experiments are run in CINECA, the Tier0 supercomputing center for scientific research in Italy [4].

The article is organized as follows: Section 1.1 presents the related works. Section 2 describes the LoRa capabilities, while Section 3 discusses the hardware design. Section 4 describes the WSN setup and the reached results. Eventually, Section V concludes the paper with comments and final remarks.

## 1.1 Related Works

Thermal management of data centers has been studied in depth in the last few years [6], ranging diverse strategies, with the goal to reduce the cooling infrastructure power consumption, improving the overall efficiency. In this context, numerous proposals have been presented: in [3] optimization applies to select a data center cooling mode and liquid flow with the aim of minimizing the facilities overall power, under the quality of service and thermal requirements. Optimization in [16] and [17] aiming to select the rack fans speed, whereas in [18], the CRACs internal temperatures are used as the reference to minimize the computational and thermal power. In [19], the effort is put into selecting the active subfloor tiles opening and blowers speed, for optimizing the Computer Room Air Handlers (CRAHs) cooling system. In [14] an optimal control policy is presented for hybrid systems featuring free, liquid and air cooling. The optimal control policy is based on a predictive model of the cooling efficiency based on environmental, room and IT temperature measurements. All model-predictive approaches implicitly assume the availability of high-quality temperature data, both off-line for model identification and online for driving control decision. In [20] a green cooling system is proposed; the management model collects information from WSN that utilizes temperature sensors to control the ventilation system and the air conditioning. The WSN is based on the ZigBee protocol and includes the actuators.

This generates a highly sophisticated network and a complex deployment even with 10 boards scattered in only 20m$^2$. As a comparison, our deployment is much cheaper because does not need any router, and at the same cost we could deploy more than 30 nodes. In our test, we covered a 150m$^2$ data center room transmitting data to a gateway installed another area of the building at a 60m distance. Moreover, some of the sensor nodes proposed in [21] require a wired power supply that severely restricts the deployment and increases the installation cost.

Another recent WSN for environmental data monitoring is presented in [5] and proposes Zigbee sensor nodes supplied by 2000mAh batteries. The deployment consists of 10 nodes in a 30m$^2$ area that represents half of their data room. Since the network is configured as a very dense mesh to provide reliability in case of packet loss, several boards are programmed as a router, and communications from the more distant node need up to 4 hops. This has a tremendous impact on the power consumption and on the lifetime of the installation. A sensor in [5] consumes 73mA on average that is two orders of magnitude higher than our LoRa solution. With an average current consumption of 194uA, our deployment can operate unattended for more than 200 days using a 1000mAh battery, half of the size needed in [5].

## 2 LoRa wireless modulation capabilities

LoRa™ is a wireless modulation for long-range low-power low-data-rate applications developed by Semtech. LoRa is supported by an alliance (LoRa Alliance) that has defined LoRaWAN, standardizing the higher-layer protocols on top of the physical radio to regulate secure communication for IoT applications and wide area networks. The network consists of end devices and gateways. Based on the LoRaWAN specifications three classes of end devices are defined: Class A, Class B, and Class C.

LoRa modulation is both bandwidth and frequency scalable. Moreover, due to the high Bandwidth Time Product (BT), a LoRa signal is very resistant to both in-band and out-of-band interference mechanisms. Since the symbol period can be longer than the typical short-duration of a noisy spike, it provides immunity to pulsed interference mechanisms. With spread spectrum, the wireless communication issues caused by the presence of interference is reduced by the process gain that is inherent to the modulation. These interfering signals are spread beyond the desired information bandwidth and can be easily removed by filtering at the receiver side.

Another factor that must be considered is the wireless link budget that defines the maximum communication range for given transmission power. The link budget delta, from a comparison between an FSK transceiver with a sensitivity of -122 dBm at 1.2Kbps with LoRa, at a fixed transmission output power, is more than four times, such as deeply studied in [7].

As well known, in a wireless path the propagation loss increases with the distance between nodes; this means that, for narrowband systems, additional nodes could be needed to generate a mesh network topology (with increased network complexity and redundancy) or to operate as additional repeaters for star networks. Unfortunately, the installation cost associated with installing a repeater increases the overall price of the WSN, in term of hardware components and software development. LoRa can minimize this cost, using a simple star network, by taking advantage of the property that signals with a different spreading factor or sequence will appear as noise at the gateway.

## 3   Network Hardware Design

The sensor node is designed to be low-power and versatile with internal temperature and humidity sensor in parallel with a thorough assessment of the components cost. Moreover, multiple sensors and devices can be connected to their expansion connectors, useful for future works. The gateway used for tests is a commercial product [22] easily customizable in hardware and software. It can be configured both as gateway alone or gateway and server.

### 3.1  Sensor Node

The end-node is based on the STM32L4 MCU from ST Microelectronics; this component includes both low-power and high computational resources, while the RFM96 SoM transceiver [23] manages the LoRa Physical layer. The high sensitivity in reception (-148 dBm) combined with the integrated +20 dBm power amplifier makes it optimal for applications requiring range or robustness in communications.

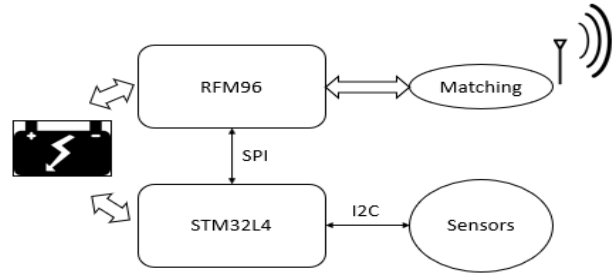

**Figure 1 -** Sensor node schematic.

The power consumption of each element was measured. The obtained current in sleep mode is 4 uA @ 3V with the Real Time Clock (RTC); instead, the STM32L4 provides a low consumption in RUN mode @48 MHz about 8.25 mA, the analog circuits (2 mA) and the RFM96 (76 mA in TX at 10 dBm) are the most expensive parts. The sensor node firmware is based on the I-CUBE-LRWAN libraries package, which is configured to be compliant with LoRaWAN Class A. Each sensor node is programmed to transmit a packet to the gateway every 30 seconds. Every packet includes a temperature and humidity sample, furthermore, to monitor and manage the WSN, node status information is acquired, such as the battery voltage and channel conflicts. With this configuration and with a battery of 1000mAh, the sensor node lifespan is seven months. Notice that the solution presented in [21] can operate only a couple of days, using the same energy stored in the battery.

### 3.2 LoRaWAN Gateway and Server

MultiConnect® Conduit$^{TM}$ is a highly configurable, manageable, and scalable communications gateway for industrial IoT applications. Network connectivity choices to any preferred data management platform include carrier approved 4G-LTE, 3G, 2G and Ethernet. A diverse range of accessory cards provide the local wired or wireless field asset connectivity and plug directly into the rear of the Conduit gateway. The LoRaWAN radio front-end includes a Semtech SX1301 and two SX1257 that demodulate the packets received simultaneously on all channels.

## 4 Network Setup

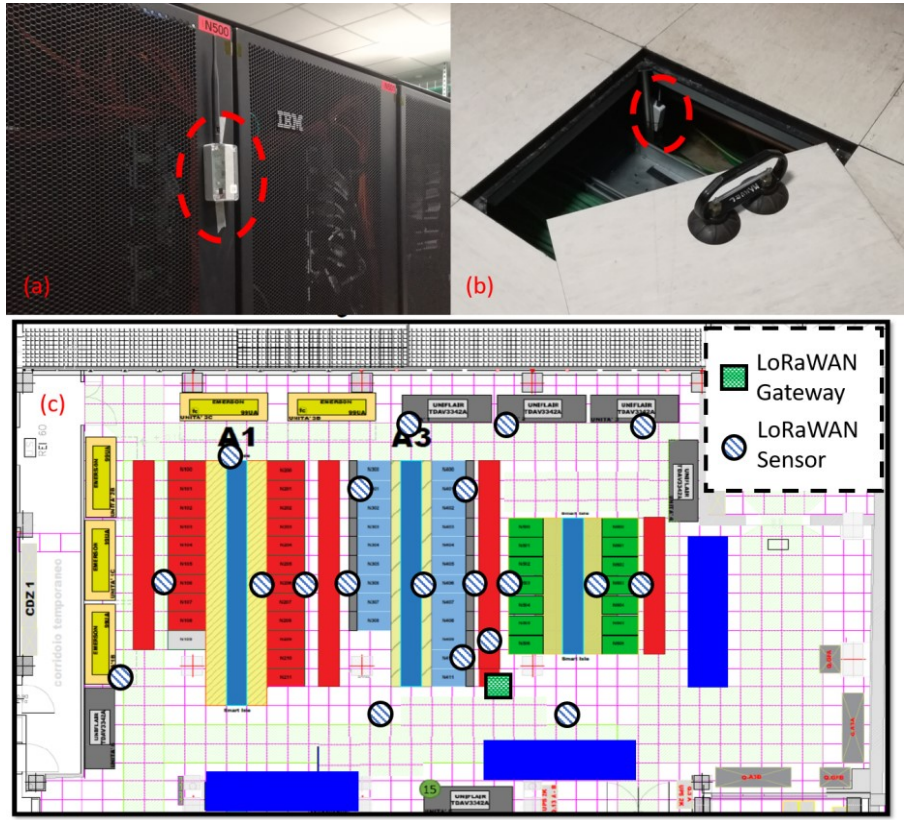

**Figure 2 -** Sensor node deployment. (a) a hallway positioning is proposed, (b) the sensor node is under the data center floor, (c) GALILEO (CINECA) data center map

A final network experimental deployment was carried out with 20 sensor nodes, arranged in key points of data center's room. This test aimed to verify in a realistic operating condition the LoRa capabilities in a noisy wireless environment. We positioned all the devices in hallways between racks, close to the CRAC output and under the floor, besides, some sensor was placed into full metallic air conditioning pipes and structure. Figure 2 shows two pictures of the WSN deployment.

### 4.1 Network results
We tested the sensor node for six months in CINECA data center, without recharging the batteries for the entire trial duration. During this time, we acquired more than six million measurements. Taking as reference a month of operation, with 1,073,771 points acquired, the LoRa radio conflicts were in average the 0.55%, that allows high reliability in term of packet communication and data analysis, a prerequisite for automatic cooling systems. By employing automatic retransmission of the collided packets, no discontinuity of collected data was detected, and only one sensor node, during the entire trial period was rebooted due to firmware issues.

All the collected data are acquired by InfuxDB time-series database, which is managed by a Node-Red application. Moreover, to provide a ready-to-use cooling monitoring system we used Grafana (Figure 3), a platform for data analytics and monitoring. This tool is used by the data center's operators to dynamically adjust the room temperature accordingly with IT's heat generation.

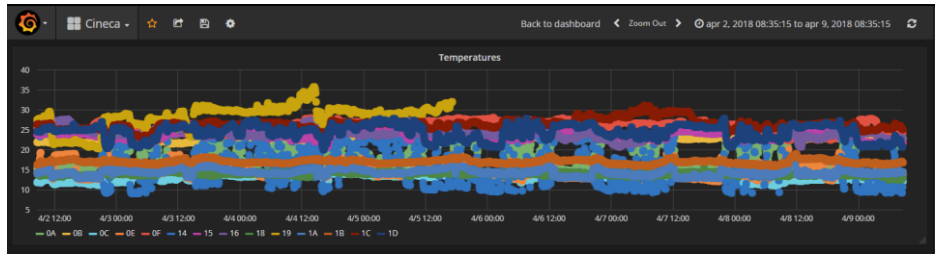

**Figure 3** - Grafana dashboard, temperatures acquired in CINECA data center.

With an average current consumption of 194uA, due to the unique characteristics of LoRa modulation, our WSN is capable to operate with a single hop communication under challenging environments, allowing both communication reliability and low power. In comparison with other solutions (Table 1), we propose a reduction of the average number of hop (ANH) up to 4 times and an improvement of 90x the battery lifetime respect ZigBee protocol and 1.5x with others LoRaWAN WSN.

Table 1 - WSN in data center environments

| WSN | Protocol | Power Supply | ANH | Max Range | Battery Lifetime |
|---|---|---|---|---|---|
| This Paper | LoRaWAN | 1000mAh Battery | 1 | 100m | 6 months |
| [20] | ZigBee | Wired | 3 | 10m | - |
| [21] | ZigBee | 2000mAh | 4 | 10m | 48 h |
| [1] | LoRaWAN | 1000mAh | 1 | 100m | 4 months |

## 5 Conclusion

This paper demonstrates that a LoRa wireless sensor nodes can be an effective and low-cost tool for temperature and humidity sensing in a data center. The proposed nodes and LoRaWAN WSN can provide reliable data for data center's room and environmental monitoring, and to train and feed numerical models for optimal cooling control. LoRaWAN network offers the advantage of easy deployment throughout the data center facilities because there is no need of wiring for power and communication. This network also offers freedom in deployment, as the sensor module can be placed in locations where wired sensors would be unfeasible for technical or safety reasons and are not constrained to keep specific distance between nodes and routers like other mesh-oriented protocols.

## Acknowledgment

This work was partially supported by a collaboration grant with CINECA. A special thanks for the support to Michele Toni, Massimo Alessio Mauri, Emanuele Sacco is also acknowledged.

## References

[1] C.L. Belady. "In the data center, power and cooling costs more than the equipment it supports" ElectronicsCooling magazine, 3(1), February 2007.
[2] M. Rossi, L. Rizzon, M. Fait, R. Passerone and D. Brunelli, "Energy Neutral Wireless Sensing for Server Farms Monitoring," in *IEEE Journal on Emerging and Selected Topics in Circuits and Systems*, vol. 4, no. 3, pp. 324-334, Sept. 2014.
[3] J. Kim, M. Ruggiero, and D. Atienza, "Free cooling-aware dynamic power management for green datacenters," Proc. of IEEE HPCS, pp. 140–146, 2012.
[4] Top500 list, Available online: https://www.top500.org/
[5] J. Dongarra. "Visit to the national university for defense technology changsha". China, University of Tennessee, 2013.
[6] R. Rhomadon, M. Ali, A. M. Mahdzir, and Y. A. Abakr, "Energy efficiency and renewable energy integration in data centers. Strategies and modeling review," Renewable and Sustainable Energy Reviews," vol. 42, pp. 429–445, 2015.
[7] Park, Seonghyun, and Janghoo Seo. "Analysis of air-side economizers in terms of cooling-energy performance in a data center considering exhaust air recirculation." Energies 11.2, 2018.
[8] C. Conficoni, A. Bartolini, A. Tilli, C. Cavazzoni and L. Benini, "HPC Cooling: A Flexible Modeling Tool for Effective Design and Management," in IEEE Transactions on Sustainable Computing.
[9] D. Porcarelli, D. Brunelli, L.Benini, "Clamp-and-Forget: A self-sustainable non-invasive wireless sensor node for smart metering applications," Microelectronics Journal, Volume 45, Issue 12, 2014, Pages 1671-1678
[10] D. Balsamo, D. Porcarelli, L. Benini and B. Davide, "A new non-invasive voltage measurement method for wireless analysis of electrical parameters and power quality," *SENSORS, 2013 IEEE*, Baltimore, MD, 2013, pp. 1-4.
[11] LoRa™ Modulation Basics, AN1200 v22, , LoRa Alliance, Inc. 2400 Camino Ramon, Suite 375 San Ramon, CA 94583 (2015)," LoRa Alliance, Tech.
[12] D. Brunelli; E. Bedeschi; M. Ferrari; F. Tinti; A. Barbaresi; L. Benini "Long-range Radio for Underground Sensors in Geothermal Energy Systems, in: Applications in Electronics Pervading Industry, Environment and Society." Lecture Notes in Electrical Engineering, vol 429. Springer, Cham, Springer (2016)
[13] D. Sartori and D. Brunelli, "A smart sensor for precision agriculture powered by microbial fuel cells," *2016 IEEE Sensors Applications Symposium (SAS)*, Catania, 2016, pp. 1-6.
[14] J. Haxhibeqiri, A. Karaagac, F. Van den Abeele, W. Joseph, I. Moerman and J. Hoebeke, "LoRa indoor coverage and performance in an industrial environment: Case study," 2017 22nd IEEE International Conference on Emerging Technologies and Factory Automation (ETFA), Limassol, 2017, pp. 1-8.
[15] É. Morin, M. Maman, R. Guizzetti and A. Duda, "Comparison of the Device Lifetime in Wireless Networks for the Internet of Things," in IEEE Access, vol. 5, pp. 7097-7114, 2017.

[16] R. Das, J. O. Kephart, J. Lenchner, and H. Hamann, "Utility function-driven energy-efficient cooling in data centers," Proc. Of ICAC, pp. 1526–1544, 2010.
[17] A. Banerjee, T. Mukherjee, G. Varsamopoulos, and S. K. Gupta, "Energy-optimal dynamic thermal management: Computation and cooling power co-optimization," IEEE Trans. Ind. Informat., vol. 6(3), pp. 340–351, 2010.
[18] L. Parolini, B. Sinopoli, B. H. Krogh, and Z. Wang, "A cyber-physical systems approach to data center modeling and control for energy efficiency," Proc. of the IEEE, vol. 100(1), pp. 255–268, 2012.
[19] R. Zhou, Z. Wang, C. E. Bash, A. McReynolds, C. Hoover, R. Shih, N. Kumari, and R. K. Sharma, "A holistic and optimal approach for data center cooling management," Proc. of IEEE Amer. Ctrl. Conf., pp. 1346–1351, 2011.
[20] Liu, Qiang, et al. "Green data center with IoT sensing and cloud-assisted smart temperature control system." Computer Networks 101 (2016): 104-112.
[21] Rodriguez, Michael G., et al. "Wireless sensor network for data-center environmental monitoring." Sensing Technology (ICST), 2011 Fifth Int. Conference on. IEEE, 2011.
[22] MultiConnect® Conduit™, programmable gateway with Linux. Available online: http://www.multitech.net/developer/products/multiconnect-conduit-platform/conduit/
[23] 868/915Mhz RF Transceiver Module. Available online: http://www.hoperf.com/rf_transceiver/lora/RFM95W.html